# A NOVEL GROUP-BASED CRYPTOSYSTEM BASED ON ELECTROMAGNETIC ROTOR MACHINE

## ASHISH KUMAR[a1] AND N S RAGHAVA[b]

[ab] **Delhi Technological University, Delhi**

## ABSTRACT

In this paper, an algorithm is aimed to make a cryptosystem for gray level images based on voice features, secret sharing scheme and electromagnetic rotor machine. Here, Shamir's secret sharing $(k, n)$ threshold scheme is used to secure a key along with voice features of $(n - k)$ users. Keystream is molded by coefficients of a voice sample, using this key stream, rotor machine's rotating cylinders' positions are initialized and internal wiring is decided by pseudo random number of Hénon chaotic map, where initial seed for chaotic system is chosen from keystream. And furthermore, shares of key stream are distributed among users. Speech processing is fused with electromagnetic machine to provide authentication as well as group based encryption. Perceptual linear predication (PLP) coefficients are utilized for formation of secret key. Simulation experiments and statistical analysis demonstrate that the proposed algorithm is sensitive to initial secret keystream, entropy, mean value analysis and histogram of the encrypted image is admirable. Hence, the proposed scheme is resistible to any vulnerable situation.

**Keywords:** Electromagnetic machine, Hénon chaotic map, PLP coefficient, secret sharing scheme

With the recent growth of multimedia, Web world is now being focused upon multimedia-based information over internet; due to this security is an important key concern while transmitting or storing information [1]. There are two types of cryptography method which is used to secure information. One of them is symmetric key cryptography [2], [3] and another one is asymmetric key cryptography. In asymmetric key cryptography scheme both sender and receiver use the different key. In traditional cryptography system, it was difficult to secure large size of multimedia from intruder or attackers and calculation of mathematical equation (built-in Encryption technique) was not so easy therefore many researchers worked upon the security of bulky information. In this series, Chaotic system, biometric features, confusion and diffusion are materialized in cryptography followed the concepts of tradition cryptography. Rotor cipher was effectively used in past; Enigma machine is an example of Rotor Machine. The Enigma machines were developed as a chain of electro-mechanical rotor cipher machines and used in the Era of mid-twentieth century to protect confidential information, diplomatic and military communication. Arthur Scherbius Enigma invented enigma machine at the end of World War II [4]- [5]. Enigma machine has a set of independent wheels through which the electric pulse can flow to other wheels. Each cylinder or wheel has fixed amount of input pins and output pins. Each input pin is connected to a unique output pin of the cylinder. So there is a unique path between input pins and output pins.

If machine has three cylinders or wheels, then it is categorized into fast rotor, medium rotor and slow rotor. Whenever any input is given, fast rotor is shifted circularly in clockwise direction and according to prior connections of wires, internal connections between pins are also shifted towards the rotor movement. A rotor with n number of labels completes its one rotation after n number of inputs. When fast rotor completes one cycle then middle rotor rotates in clockwise direction by unit position. Slow rotor shifts one position to right after one complete cycle of middle rotor. This movement makes the system dynamic in nature.

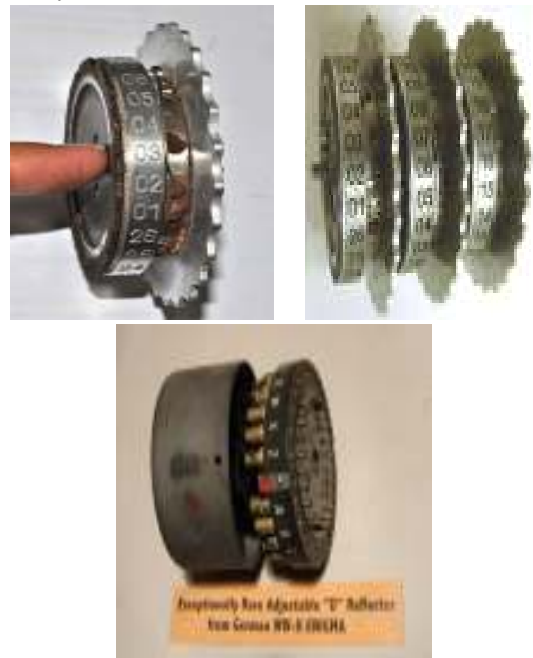

**Fig. (a) :War-damaged Enigma rotor A7135 (b) Three rotors on their shaft (c) Original Enigma "D" Reflector number A5221**

[1]Corresponding Author

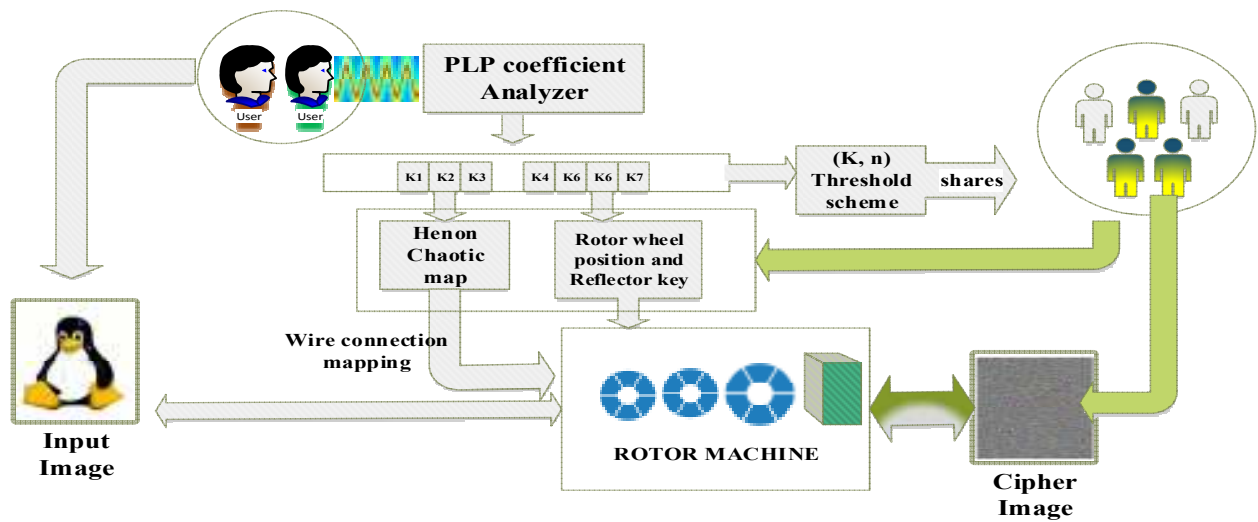

Fig.2: The architecture of Proposed Cryptosystem

Every person has their own way to speak a language, this voice sample is preprocessed and further synthesized for various applications such authentication. When any person speaks, the sound is perceived by human ear or machine. MFCCs (Mel Frequency Cepstral Coefficients) and PLP (perceptual linear predication) coefficients of Voice are unique biometric feature which can be used to provide security services according to security requirements [8]- [9]. In PLP perceptual property of human ear is captured. Power spectrum of speech signal in bark scale is equivalent to human's perceptual model. MFCC coefficients are find out under the Mel scale filters which are triangular filters where as PLP coefficients are find out under Bark scale filters which are trapezoidal in shape. In cryptography speech processing is being used with other biometrics at the vast level for reliable security systems.

Related Work

Zhi-liang ZHU at el. proposed an encryption method based on nonlinear mechanism of enigma machine and chaos controlling the encryption process [10]. In paper [11], Elkamchouchi and Elshafee proposed a rotor enhanced block cipher method to obtain permutation and substitution operations in which rotor generates the round key to achieve cipher text key dependency. In paper [12], author proposed cryptographic system with unbalanced rotor to achieve goals of permutation and substitution.

In [13], Chen at el. presented a symmetric image encryption scheme based on 3D chaotic cat maps. Wang at el. [14] introduced a 3D Cat map based symmetric image encryption method which comes out to be computationally expensive process and has a non-independent key space. Image encryption algorithms are constructed on diffusion property of shannon's theory where hyper chaotic systems are also presented in cryptosystem [15], [16]. In [21], Wahyudi, Astuti, and Syazilawati adopted PLP coefficients of voice to develop intelligent voice-based door access control system.

**PROPOSED METHOD**

Proposed algorithm is designed having regard to sensitive information of Digital images. Algorithm is performed within two groups where One group encrypt the input image with their voice segment and decomposed secret key into shares and transmitted to other group. Therefore, in the first phase, key generation and encryption process take place, where form group $A$, a person speaks from sound acquisition device and this wave sample converted into keystream for the further process and after that designing process of rotor machine is done by Hénon chaotic system. Here cryptosystem is completely based on the concept electromagnetic rotor machine and properties of Hénon chaotic map. Rotor machine is vital core part of the second phase of algorithm.

Let us consider digital color image $I$ with dimension. $M \times N \times 3$ (i.e. M rows and N columns). Digital color image is transformed into gray scale image. This gray scale image supports $[0\ 255]$ decimal value with 8-bit binary format of computer system.

**PLP (Perceptual Linear Prediction) Feature Through Voice Segment**

Step 1: choose speech acquisition tool to capture voice segment and take 256 sample from this wave file.
Step 2: calculate FFT and calculate power spectrum using squared magnitude of signal.

Step 3: covert frequency bin point to corresponding bark and bark scale is divided into equal 14 filters.

Step 4: calculate cube root of the power spectrum for each bark scale values.
Step 5: Bark scale is divided equally spaced triangular filters, width of filters equal to 5 bark scales with 50% overlap.
Step 6: Calculate IFFT of each triangular filter, these 14 values are known as PLP coefficients.

Step 7: out of 14 coefficients, 7 coefficients are chosen by the person to generate a keystream [17]- [18] and following steps which are given below:

$$keystream \leftarrow K[7]$$
$$|K[7]| \leftarrow K[7] \times 256$$
$$K[7] \leftarrow MOD(K[7], 256)$$
$$Keystream \leftarrow K[7]$$

**Digital Electromagnetic Rotor Machine**

Electromagnetic machines used in World War II to send secret information, where each rotor's pins were labeled with 26 characters and mapping of wires within rotor was static all the time. So basically it was based on the initial position of the rotor and the wired mapping of reflector. Here we have designed a cryptosystem which is truly based on Electromagnetic machine concept. It is a digital model of electromagnetic machine to encrypt information with more number of combinations and dynamic wired mapping between pins, so it is hard to eavesdropper to deduce the original content from cipher. Here we have designed rotor with 256 input pins and each pin is connected with their corresponding output pins. Output pins label are generated using Hénon chaotic map and according to that, input pins are connected to output pins with their corresponding number. For example, a rotor has 4 pins. Initial rotor position starts from $K1 = 4$ and $K4$ be some number as initial seeds for Hénon chaotic map which generates a sequence, say, [ 3 2 1 4 ].

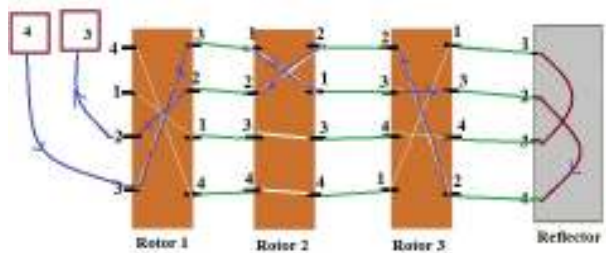

**Fig.3: Wire connections within Rotor**

Figure 3 illustrates the wiring connections among rotors and reflector. For the given example initial input at 4[th] pin gives output at 3[rd] pin. Output pins label of rotor 1 are generated using Hénon chaotic map and according to that, input pins are connected to output pins with their corresponding number. Blue path shows the flow of current within the model with respect to the input.

**A. Algorithm for Encryption**

Step 1: $K1, K2, K3$ are the initial seeds for Hénon chaotic map using (1), which generates the sequence for Rotor1 Rotor2 and Rotor3 respectively, as shown in fig. 2.
Step 2: keystream $K4, K5, K6$ are used to initialise the positions of all rotors in the machine. $K7$ is used to connect pins within reflector.

Step 3: Hénon chaotic map [19] discovered in 1978 is used as a pseudo random number generator in security systems. Two dimensional discrete-time nonlinear dynamical Hénon chaotic map generates pseudo-random binary sequence which has been described as below:

$$X_{n+1} = 1 + Y_n - aX_n^2$$
$$Y_{n+1} = bX_n \text{ where } n = 0, 1, 2 \ldots \quad (1)$$

Here, the parameters, *a* and *b* are of prime importance as the dynamic behavior of system depends on these values. The system cannot be chaotic unless the value of a and b are 1.4 and 0.3 respectively.

Step 4: This sequence is converted into $[1\ 256]$ range using modular arithmetic.

$$X \in \{-I, +I\} ; \text{ Where } I \text{ is integer}$$
$$New\_X \leftarrow floor\{(256 \times X_{256}\}$$
$$New\_X \leftarrow MOD(New\_X, 256)$$

Step 5: remove duplicate numbers from the sequence and replace by those numbers with 0 frequency count in sequence list as shown in fig. 4.

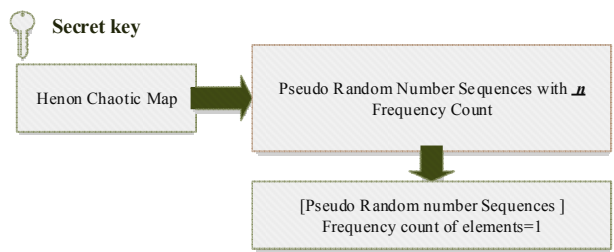

**Fig.4: Pseudo random numbers for wire connections within Rotor**

Step 7: Using the pseudo random sequence, labeling of output pins are done and wire are mapped with their corresponding label. This wiring of the model remains same until next setup.

Step 8: grayscale image is used as an input, where each pixel intensity is responsible for movement of fast rotor. Every pixel goes as input one by one into the system and gives output calculated by the below mentioned pseudo code for encryption of image.

```
For I ← 1:M
   For J ← 1: N
   Ptr ← I (I, j) +1
   For Rotor=1:3
      Number ← rotor (input pin, ptr)
      search (number1, rotor1) in output pin
      return index where number is found
   end
   Ref_indx ← mod(index[rotor3]+K7),256)
   For Rotor ← 3:1
      search (number1, rotor) in input pin
      return index where number is found
   end
   end
end
```

**Key Generation for Group B Members**

Secret sharing scheme [20] basically splits the secret into shares and these shares are further distributed to the concerned group. Secret can be constructed only when group members participate to generate secret key. Here Shamir's secret sharing *(k, n)* threshold scheme is used to split secret n members of the group. Therefore, parameters are taken as *n=5, K=3 and P=17*.

An equation is created with degree *(K-1)*, and the coefficients ($a_1, a_2$) of the equation are chosen by group *A* members. Keystream is processed in secret sharing scheme and *n* shares of keystream's size are distributed to the concern group. Where secret keystream is written in place of $a_0$.

$$f(x) = a_0 + a_1 x + a_2 x^2 \quad (2)$$

**B. Decryption procedure**

Group member can reconstruct a secret key using LaGrange interpolation method which is given below:

$$p(x) = \sum_{j=1}^{n} P_j(x)$$

$$\text{Where, } P_j(x) = Y_j \prod_{\substack{k=1 \\ k \neq j}}^{n} \frac{x - x_k}{x_j - x_k} \quad (3)$$

Step 3: At decryption end, users can generate keystream using (3) to reconstruct original image using the rotor machine.

**EXPERIMENTAL RESULTS**

In this section, experimental results of the proposed image encryption algorithm are given to appreciate the efficiency of proposed security system. MATLAB 7.9 software is used for implementation of proposed algorithm. Figure 11 shows PLP coefficients of voice sample (.wav format) used to generate keystream. Results of proposed system are shown in Fig.5-10 and table I.

**Encryption process**

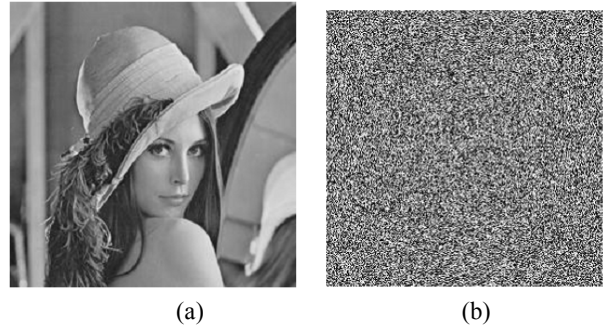

(a)       (b)

Fig.5.:Lena: (a) original image; (b) cipher image.

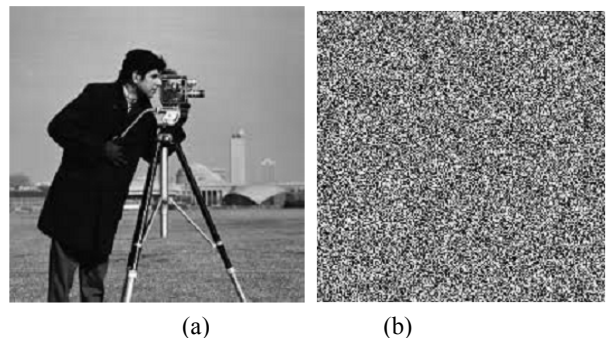

(a)       (b)

Fig.6:Cameraman: (a) original image; (b) cipher image

**Decryption Process**

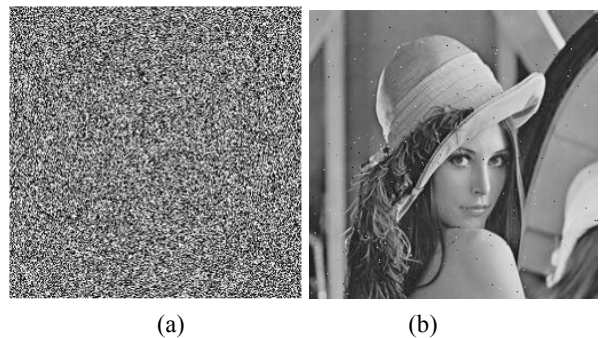

(a)       (b)

Fig.7:Lena: (a) cipher image; (b) decrypted image

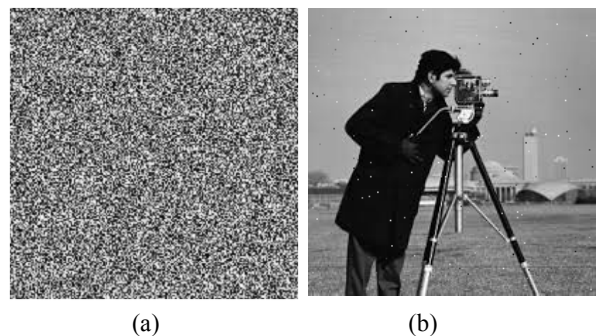

(a)       (b)

Fig.8: Cameraman: (a) cipher image; (b) decrypted image

**Entropy Analysis of Results**

Entropy of encrypted image decides ability of a cryptosystem which makes it difficult for eavesdropper to deduce information from cipher image. Ideal entropy of a cryptosystem is $\cong 8$ which means uniform distribution of pixel values.

**Histogram**

Figure 9 shows uniform distribution of gray scale pixel values in cipher image, and significantly different from histogram of original image which proves that encrypted image does not help intruders to employ statistical attack on encryption procedure.

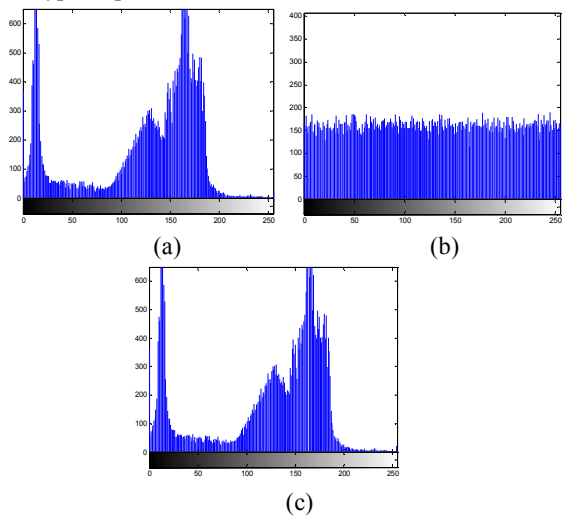

Fig.9.:Cameraman: (a) histogram of original image; (b) histogram of cipher image; (c) histogram of decrypted image

**Table I: Entropy Analysis**

| Image name | Entropy of original image | Entropy of encrypted image |
|---|---|---|
| Cameraman.jpeg | 7.086010767836325 | 7.996133299611271 |
| Lena.jpg | 7.466871535624320 | 7.997155374413349 |

**Mean Value Analysis**

Mean value analysis gives average intensity of pixels in horizontal direction across the image. Mean value of cipher image in fig. 10 is shown by green color, which is consistent throughout. This indicates uniform distribution of gray levels whereas decrypted image and original image are shown by red and blue color respectively. Both the lines overlap each other which means original image is obtained by the group after decryption.

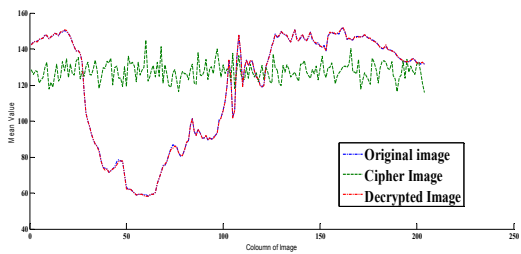

Fig.10: Cameraman: (a) original image; (b) cipher image

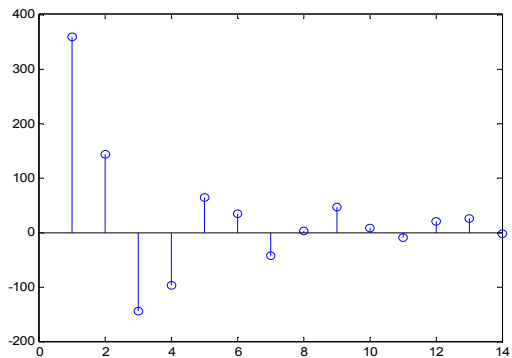

Fig.11. PLP coefficients of voice sample

**CONCLUSION AND DISCUSSION**

In this paper, property of electromagnetic machine is used with a novel approach to achieve confidentiality and authentication with high security. Speech is used as for authentication process and substitution cipher is achieved by electromagnetic machine which is based on the keystream and Hénon chaotic map. A system with three rotors will have $256^3$ different combinations and same is followed in mapping of wires within rotor. Since chaos systems are very sensitive to initial condition so a slight change in initial key gives a different result and makes it impossible for intruder to break the cipher image. The proposed cryptosystem is applied on several test image and results show a high level of security given by system. For few test images, the decrypted images were found to have insignificant noise. Here, the security of system also relies on a speech signal along with the electromagnetic machine.